\documentclass[conference,10pt]{IEEEtran}
\usepackage{graphicx}
\usepackage{amsmath}
\usepackage{amssymb}
\usepackage{multirow}

\usepackage{cite}
\usepackage{color}
\usepackage{amsthm}

\usepackage{algorithm}
\usepackage{algpseudocode}
\usepackage{pifont}
\usepackage{cleveref}
\usepackage{epstopdf}
\usepackage{url}
\newcommand{\mat}{\mathbf}

\renewcommand{\vec}[1]{\bar{\mat{#1}}}
\graphicspath{{./figures/}}

\begin{document}

\title{Optimal Harvest-or-Transmit Strategy for Energy Harvesting Underlay Cognitive Radio Network}


\author{\IEEEauthorblockN{Kalpant Pathak and Adrish Banerjee}
\IEEEauthorblockA{Department of Electrical Engineering, Indian Institute of Technology Kanpur, Uttar Pradesh, India 208016\\
E-mail: \{kalpant, adrish\}@iitk.ac.in}}

\maketitle
\vspace{-1cm}
\begin{abstract}
An underlay cognitive radio network with energy harvesting is considered which operates in slotted fashion. The primary user (PU) transmits with a constant power in each slot, while the secondary user (SU) either harvests energy from primary's transmission or transmits its data. We propose an optimal offline \emph{harvest-or-transmit} strategy where in each slot, SU takes a decision whether to harvest energy or transmit its data limiting interference at the primary receiver. We aim to maximize the achievable rate of SU under energy causality and interference constraints. The optimization problem is formulated as a mixed integer non-linear program and the optimal \textit{harvest-or-transmit} policy is obtained using \emph{generalized Benders decomposition} algorithm. Through simulations, we analyze the effects of various system parameters and interference constraint at the primary receiver on the optimal policy.
\end{abstract}

\IEEEpeerreviewmaketitle

\section{Introduction}
In a wireless communication system, two major challenges are to achieve high spectral and energy efficiency. One of the possible solution for these two challenges is energy harvesting cognitive radio network (EH-CRN) \cite{eh_crn_survey}. In EH-CRNs, a set of users namely licensed (primary) and unlicensed (secondary) users (PU and SU respectively) share the same spectrum while harvesting energy from the environment. EH-CRNs have been studied operating in interweave mode in \cite{interweave_1,interweave_2,interweave_3,interweave_4}, overlay mode in \cite{overlay_2,overlay_3,overlay_4,overlay_5} and underlay mode in \cite{underlay_1,underlay_2,underlay_3}. 

In underlay CRNs, the PU and SU users coexist and the SU transmits along with PU while limiting the interference at primary receiver (PR). In \cite{underlay_1}, an underlay EH-CRN is considered where SU harvests energy at the beginning of each slot. The authors used geometric waterfilling with peak power constraint (GWFPP) to obtain an optimal offline power allocation policy for SU which maximizes the throughput. In \cite{underlay_2}, the cooperation between energy harvesting PU and SU is considered at energy level. In each slot, SU may transfer some fraction of its energy to PU and transmits along with it. Authors obtained transmission policies maximizing SU's throughput and showed that energy cooperation helps secondary improve its performance. In \cite{underlay_3} and \cite{my_spcom}, authors considered a scenario where in each slot, the SU harvests energy from PU's transmission for some fraction of the slot and transmits its data in the remaining fraction. The authors obtained a suboptimal myopic transmission policy in \cite{underlay_3} and an optimal offline  transmission policy in \cite{my_spcom} maximizing SU's achievable throughput under outage constraint of PU.

We consider the system model similar to \cite{underlay_3} and \cite{my_spcom}. However in our model, each slot is dedicated either for energy harvesting or information transfer (\textit{harvest-or-transmit} policy). This policy makes the switching between the harvesting module and transmission module less complex by allowing less frequent switching ($N-1$ switchings in worst case as compared to $2N-1$ in \cite{underlay_3} and \cite{my_spcom} for $N$ slots). In addition, unlike the time sharing policy, switching occurs only at the end of the slot which results in less complex switching circuitry. We are interested in finding an optimal offline \textit{harvest-or-transmit} policy which acts as a benchmark for the online and suboptimal offline policies for the system model under consideration, and gives an upper bound on the system performance. The channel gains can be obtained using any channel prediction technique \cite{chennal_prediction}. Our contributions in this paper are as follows:
\begin{itemize}
\item We formulate the optimization problem of maximizing the achievable rate of SU over a finite number of slots under PU's interference constraint and SU's energy causality constraint as a non-convex mixed integer non-linear program (MINLP). Then, we convert the non-convex MINLP into an equivalent convex MINLP and obtained the optimal \textit{harvest-or-transmit} policy using generalized Bender's decomposition (GBD) algorithm.
\item We than analyze the effects of various system parameters on the optimal \textit{harvest-or-transmit} strategy through simulations.
\item   Finally, we compare the optimal policy with the myopic policy proposed in \cite{underlay_3}, and show that the former outperforms the latter in terms of achievable rate.
\end{itemize}
\section{System Model}
\vspace{-4mm}
\begin{figure}[!ht]
\centering
\includegraphics[width=0.65\linewidth]{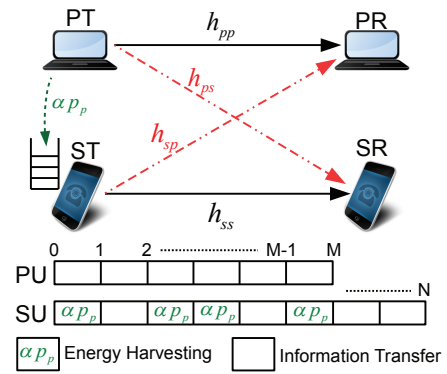}
\vspace{-2mm}
\caption{An underlay EH-CRN with \textit{harvest-or-transmit} strategy.}
\label{fig:system_model}
\vspace{-2mm}
\end{figure}
The system model is shown in Fig. \ref{fig:system_model}. The primary transmitter (PT) transmits with power $p_p$ in all the slots and remains active for total $M$ slots. This information of PU's availability is not needed to be known at the SU and in this case, the SU will follow a policy assuming that the interference constraint of PU is needed to be satisfied in all the slots. However, if it is known, the SU can optimize its transmission strategy which will improve its throughput. In either case, as long as the PU is present, in each slot, the secondary transmitter (ST) decide either to harvest energy from primary's transmission or transmit its data with power $p_s^i$ in the $i$th slot. The ST is equipped with an infinite capacity battery to store the harvested energy. The PT and ST are assumed to be in close vicinity so that the effects of multipath fading on harvested energy can be neglected. We consider a case where the ST operates for $N>M$ slots and $M$ is known, and therefore it can either harvest or transmit in the first $M$ slots with efficiency $0\leq\alpha\leq1$, and transmits without harvesting in the remaining $N-M$ slots. The ST has an interference constraint $P_{int}$ in each slot. The algorithm can also be modified for $N=M$ and $N<M$, and the extension to these cases is straightforward. We assume that the battery at the ST has an initial energy of $E_0$. We assume quasi static Rayleigh fading channel. Therefore, the power gains of all the channel links are i.i.d. exponentially distributed. We assume the slot length $\tau$ to be 1 second so that terms power and energy can be used interchangeably. However, the proposed policy can be modified for any value of $\tau$.
\section{Problem Formulation}
\indent In the system model considered, in each slot ST decides whether to harvest energy from primary's transmission or communicate with secondary receiver (SR) with an optimal power. We aim to maximize the achievable rate of ST over all slots under energy availability constraint and interference constraint ($P_{int}$) of ST and  PR respectively.\\
\indent Let us take an indicator function $I_H^i$ such that it takes value 1 if ST harvests energy in the $i$th slot and takes value 0 otherwise.
 Whenever $I_H^i=0$, ST transmits with power $p^i_s$ in $i$th slot and its instantaneous achievable rate is given by Shannon's capacity formula $R_i=\log_2\left(1+\frac{p_s^ih_{ss}^i}{\sigma^2+h_{ps}^ip_p^i}\right)$ bps/Hz for the first $M$ slots. And in remaining slots, since PU is absent, the instantaneous achievable rate of ST is given as $R_i=\log_2\left(1+\frac{h_{ss}^ip_s^i}{\sigma^2}\right)$ bps/Hz, $i=M+1,\ldots,N$, where $p_p^i=p_p$ and $p_s^i$ are the transmit powers of PT and ST in $i$th slot respectively, $h_{ss}^i,h_{ps}^i,h_{sp}^i$ and $h_{ss}^i$ are the i.i.d. exponentially distributed power gains of PT-PR, PT-SR, ST-PR and ST-SR channel link respectively, and $\sigma^2$ is the variance of the additive noise at both the receivers, which is assumed to be zero mean Gaussian (AWGN).\\
\indent The optimization problem ($\textbf{P}_1$) of maximizing the achievable rate of ST under energy causality constraints and interference constraint of ST and PR respectively, is written as:
\begingroup 
\allowdisplaybreaks
\begin{subequations}\label{original_opt}
\begin{align}
\max_{\small{\begin{array}{c}
\vec{p}_s\succeq0\\
\vec{I}_H\in \{0,1\}^M
\end{array}}} & f(\vec{p}_s,\vec{I}_H) \label{eq:opt_1}\\
\text{s.t.\hspace{3mm}}\quad & (1-I_H^1) p_s^1\leq E_0,\label{eq:opt_2}\\
& (\text{Energy causality constraint for 1$^{st}$ slot})\nonumber\\
& \hspace{-1.8cm}\sum_{j=1}^i(1-I_H^j) p_s^j\leq E_0+\alpha\sum_{j=1}^{i-1}I_H^j  p_p^j,\,i=2,\ldots,M, \label{eq:opt_3}\\
&\!\!\!\hspace{-8mm}(\text{Energy causality constraint for 2$^{nd}$ to $M^{th}$ slot})\nonumber\\
&\!\!\!\!\!\!\!\hspace{-8mm} \sum_{j=1}^M(1-I_H^j) p_s^j+\sum_{j=M+1}^i p_s^j\leq E_0+\alpha\sum_{j=1}^{M}I_H^j p_p^j,\nonumber\\
& \quad \quad \quad \quad \quad \quad \quad  i=M+1,\ldots,N,\label{eq:opt_4}\\
&\!\!\!\hspace{-7mm}(\text{Energy causality constraint for remaining slots})\nonumber\\
&\!\!(1-I_H^i)h_{sp}^ip_s^i\leq P_{int},\quad i=1,\ldots,M,\label{eq:opt_5}\\
& \quad\quad\quad\quad(\text{Interference constraint at PR})\nonumber
\end{align}
\end{subequations}
\endgroup 
where $f(\vec{p}_s,\vec{I}_H)=\sum_{i=1}^M (1-I_H^i)\log_2\left(1+\frac{h_{ss}^ip_s^i}{\sigma^2+h_{ps}^ip_p^i}\right)+\sum_{i=M+1}^N  \log_2\left(1+\frac{h_{ss}^ip_s^i}{\sigma^2}\right)$, $0\leq\alpha\leq1$ is the energy harvesting efficiency, $E_0$ is the initial energy available at ST, and $P_{int}$ is the acceptable interference threshold of primary receiver. Vectors $\vec{p}_s$ and $\vec{I}_H$ are such that $[\vec{p}_s]_i=p_s^i$ and $[\vec{I}_H]_i=I_H^i$, and $\vec{p}_s\succeq0$ means $p_s^i\geq0,\,\forall i$. The constraints (\ref{eq:opt_2})-(\ref{eq:opt_4}) mean that we can use only that much energy which we have harvested upto that slot.\\
\indent The problem $\textbf{P}_1$ is a non-convex MINLP as variables $\vec{I}_H$ and $\vec{p}_s$ appear in product form. However, it can be converted into convex MINLP and solved optimally using GBD algorithm \cite{GBD}.
\subsection*{Convex MINLP}
After some manipulations in the constraints, the equivalent convex MINLP $\textbf{P}_2$ of optimization problem $\textbf{P}_1$ is given as:
\begin{subequations}\label{convex}
\begin{align}
\max_{\small{\begin{array}{c}
\vec{p}_s\succeq0\\
\vec{I}_H\in \{0,1\}^M
\end{array}}}& f(\vec{p}_s)\label{eq:opt_convex_1}\\
\text{s.t.\hspace{3mm}} \quad &  p_s^1\leq (1-I_H^1)E_0,\label{eq:opt_convex_2}\\
&\hspace{-9mm}  p_s^i\leq (1-I_H^i)\left(E_0+\sum_{j=1}^{M} p_p^j\right),\quad i=2,\ldots,M,\label{eq:opt_convex_3}\\
& \hspace{-8mm}\sum_{j=1}^i p_s^j\leq E_0+\sum_{j=1}^{i-1} \alpha I_H^j p_p^j, \quad i=2,\ldots,M,\label{eq:opt_convex_4}\\
&\!\!\!\!\!\!\hspace{-8mm} \sum_{j=1}^{M+i} p_s^j\leq E_0+\sum_{j=1}^M\alpha I_H^j p_p^j, \quad i=1,\ldots,N-M,\label{eq:opt_convex_5}\\
& h_{sp}^ip_s^i\leq P_{int},\quad\quad\quad\quad i=1,\ldots,M,\label{eq:opt_convex_6}
\end{align}
\end{subequations}
where $f(\vec{p}_s)=\sum_{i=1}^M\log_2\left( 1+\frac{h_{ss}^ip_s^i}{\sigma^2+h_{ps}^ip_p^i}\right)+\sum_{i=M+1}^N \log_2\left(1+\frac{h_{ss}^ip_s^i}{\sigma^2}\right)$. The equivalence between \eqref{original_opt} and \eqref{convex} can be understood as follows. When $I_H^i=1$ for some $i\leq M$, the constraints \eqref{eq:opt_convex_2} or \eqref{eq:opt_convex_3} results in $p_s^i\leq0$, which along with constraint $p_s^i\geq0$ results in $p_s^i=0$. In this case, constraint \eqref{eq:opt_convex_4} would consider only those $p_s^i$'s which are positive. On the other hand when $I_H^i=1$ for some $i\leq M$, the constraint \eqref{eq:opt_convex_3} given an outer bound on $p_s^i$ and hence, has no effect. In this case, the constraints \eqref{eq:opt_convex_4} and \eqref{eq:opt_convex_5} will dominate and represent the energy causality constraints in \eqref{eq:opt_4} and \eqref{eq:opt_5}.

 The problem (\ref{convex}) is a convex MINLP problem since the objective function is concave in $\vec{p}_s$ and the constraints contain affine inequalities. Since the continuous variable $\vec{p}_s$ and the integer variable $\vec{I}_H$ are now linearly separable, this problem can be solved efficiently using GBD algorithm \cite{GBD}.

\begin{figure*}[!t]
\begin{multline}
\mathcal{L}(\vec{p}_s,\theta,\vec{\pmb{\lambda}}, \vec{\pmb{\gamma}},\vec{\pmb{\delta}},\mu)= f(\vec{p}_s)+\theta\left[(1-I_H^1)E_0- p_s^1\right]+\sum_{i=1}^M\mu_i[P_{int}-h_{sp}^ip_s^i]
+\sum_{j=1}^{M-1}\lambda_j\left[(1-I_H^{j+1})\left\{\sum_{i=1}^M p_p^i+E_0\right\}- p_s^{j+1}\right]\\+\sum_{j=1}^{M-1}\gamma_j\left[E_0+\sum_{i=1}^j\alpha I_H^i p_p^i-\sum_{i=1}^{j+1} p_s^i\right]
+\sum_{j=1}^{N-M}\delta_j\left[E_0+\sum_{i=1}^M\alpha I_H^i p_p^i-\sum_{i=1}^{M+j} p_s^i\right].
\label{eq:primal_lagrangian}
\end{multline}
\noindent\rule{\linewidth}{0.4pt}
\vspace{-8mm}
\end{figure*}
\section{Optimal Harvest-or-Transmit Strategy using GBD Algorithm}
\indent The GBD algorithm decomposes the problem \eqref{convex} into two subproblems: a primal and a master problem, and solves it iteratively. It solves the primal problem and gives a solution $\vec{p}_s$ for a fixed $\vec{I}_H$, which is obtained from previous iteration of master problem. This solution of master problem gives $\vec{I}_H$ for previously obtained $\vec{p}_s$ along with the corresponding Lagrange multipliers. The algorithm is initialized by random selection of $\vec{I}_H$ from the set $\{0,1\}^M$. The primal and the master problem for $l$th iteration are given as follows:
\vspace{-0mm}
\subsection{Primal Problem ($l$th iteration)}
In iteration $(l-1)$, we obtain an optimal $\vec{I}_H^{(l-1)*}$ from $(l-1)^{th}$ iteration. Then, the primal problem in iteration $l$ is given as:
\begingroup
\allowdisplaybreaks
\begin{align}
\max_{\vec{p}_s\succeq0}\quad f(\vec{p}_s),\;\;
\text{s.t.}\quad \text{(\ref{eq:opt_convex_2})-(\ref{eq:opt_convex_6})}. \label{eq:primal_1}
\end{align}
\endgroup
The optimization problem (\ref{eq:primal_1}) is convex in $\vec{p}_s$ \cite{cvx_book} and can be solved using CVX \cite{cvx}. The solution of the primal problem in $l$th iteration, $\vec{p}^{(l)*}_s$ is used to obtain the solution of the master problem in next iterate, $\vec{I}^{(l)*}_H$ for fixed $\vec{p}_s$ and dual variables $\theta,\vec{\pmb{\lambda}},\vec{\pmb{\gamma}},\vec{\pmb{\delta}}$ and $\vec{\pmb{\mu}}$ associated with constraints (\ref{eq:opt_convex_2}), (\ref{eq:opt_convex_3}), (\ref{eq:opt_convex_4}), (\ref{eq:opt_convex_5}) and (\ref{eq:opt_convex_6}) respectively.

The Lagrangian of the primal problem is given in (\ref{eq:primal_lagrangian}). The Karush-Kuhn-Tucker (KKT) stationarity conditions are:
\begin{align*}
&\Psi_1-\mu^*_1h_{sp}^1-\sum_{j=1}^{M-1}\gamma^*_j-\sum_{j=1}^{N-M}\delta^*_j=0,\\
&\Psi_i -\mu^*_ih_{sp}^i-\lambda^*_i-\sum_{j=i-1}^{M-1}\gamma^*_j-\sum_{j=1}^{N-M}\delta^*_j=0,\,\forall i\backslash\{1\},\\
&\frac{ h_{ss}^i}{\sigma^2+h_{ss}^ip_s^{i*}}-\sum_{j=i-M}^{N-M} \delta^*_j=0,\;\;\text{for } i=M+1,\ldots,N,
\end{align*}
where $\Psi_i = \frac{ h_{ss}^i}{\sigma^2+h_{ps}^ip_p^i+h_{ss}^ip_s^{i*}},\,\forall i$. The complementary slackness conditions are
\begingroup
\allowdisplaybreaks
\begin{align*}
\theta^*\left[ p_s^{1*}-(1-I_H^1)E_0\right]=&0,\\
\mu^*_i\left[h_{sp}^ip_s^{i*}-P_{int}\right]=&0,\;\;i=1,\ldots,M,\\
\lambda^*_i\left[ p_s^{i+1*}-(1-I_H^{i+1*})\zeta\right]=&0,\;\;i=1,\ldots,M-1,\\
\gamma^*_i\left[\sum_{j=1}^{i+1} p_s^{i*}-E_0-\sum_{j=1}^i\alpha I_H^{i*} p_p^i\right]=&0,\;\;i=1,\ldots,M-1,\\
\delta_i\left[\sum_{j=1}^{M+i} p_s^{i*}-E_0-\sum_{j=1}^M\alpha I_H^{i*} p_p^{i*}\right]=&0,\;\; i=1,\ldots,N-M,
\end{align*}
\endgroup
where $\zeta=\left\{\sum_{j=1}^M p_p^i+E_0\right\}$. The dual variables associated with non-negativity constraints can be neglected for mathematical ease. However these constraints can be included later by projection onto the positive orthant. Using the KKT conditions, the optimal transmit power in $l$th iteration is given as:
\begin{align}
p_s^{i*(l)}=\left\{\hspace{-2mm}
\begin{array}{ll}
\left[\frac{1}{\zeta_1}-\frac{\sigma^2}{h_{ss}^1}-\frac{h_{ps}^1p_p^1}{h_{ss}^1}\right]^{+(l)}, &\hspace{-3mm} i=1,\\
\left[\frac{1}{\zeta_i}-\frac{\sigma^2}{h_{ss}^i}-\frac{h_{ps}^ip_p^i}{h_{ss}^i}\right]^{+(l)}, &\hspace{-3mm} i=2,\ldots,M,\\
\left[\frac{1}{\sum_{j=i-M}^{N-M}\delta^*_j}-\frac{\sigma^2}{h^i_{ss}}\right]^{+(l)}, &\hspace{-3mm} i=M+1,\ldots,N,
\end{array}
\right.
\end{align}
where $\zeta_1=\theta^*+\mu_1^*h_{sp}^1+\sum_{j=1}^{M-1}\gamma^*_j+\sum_{j=1}^{N-M}\delta_j^*$, $\zeta_i=\mu_i^*h_{sp}^i+\lambda^*_{i-1}+\sum_{j=i-1}^{M-1}\gamma_j^*+\sum_{j=1}^{N-M}\delta_j^*,\, i=2,\ldots,M$, and $[x]^+$ represents $\max\{x,0\}$. The optimal primal and dual variables in $l$th iterations are obtained using CVX \cite{cvx}. The master problem for $l$th iteration is explained in next subsection.
\subsection{Master Problem ($l$th iteration)}
We require the Lagrangian of the primal problem for the formulation of master problem, which is given in (\ref{eq:primal_lagrangian}) on the top of the next page, where $\theta,\vec{\pmb{\lambda}},\vec{\pmb{\gamma}},\vec{\pmb{\delta}}$ and $\vec{\pmb{\mu}}$ are Lagrange multipliers for constraints (\ref{eq:opt_convex_2}), (\ref{eq:opt_convex_3}), (\ref{eq:opt_convex_4}), (\ref{eq:opt_convex_5}) and (\ref{eq:opt_convex_6}) respectively. Let $\theta^*,\vec{\pmb{\lambda}}^*,\vec{\pmb{\gamma}}^*,\vec{\pmb{\delta}}^*$ and $\vec{\pmb{\mu}}^*$ be the optimal Lagrangian variables. For given $\vec{p}_s^{(l)*},\theta^{(l)*},\vec{\pmb{\lambda}}^{(l)*},\vec{\pmb{\gamma}}^{(l)*},\vec{\pmb{\delta}}^{(l)*}$ and $\vec{\pmb{\mu}}^{(l)*}$ obtained from primal problem in $l$th iteration, we formulate the master problem as:
\begin{subequations}\label{eq:master}
\begin{align}
\max_{t\geq0,\vec{I}_H\in\{0,1\}^M} \quad & t\label{eq:master_1}\\
\text{s.t.\hspace{7mm}}\quad &\!\!\! t \leq \mathcal{L}\left(\vec{p}_s^{(j)*},\theta^{(j)*},\vec{\pmb{\lambda}}^{(j)*}, \vec{\pmb{\gamma}}^{(j)*},\vec{\pmb{\delta}}^{(j)*},\mu^{(j)*}\right),\nonumber\\
& \hspace{1.1in} j\in\{1,2,\ldots,l\}. \label{eq:master_2}
\end{align}
\end{subequations}
The problem (\ref{eq:master}) is a mixed integer linear program (MILP) of $t$ and $\vec{I}_H$ and hence, can be solved optimally using MOSEK \cite{mosek}.

\textbf{GBD Algorithm:} In the $l$th iteration, the master problem gives a solution $t$, an upper bound to the solution of original problem $\textbf{P}_1$. Also, in each iteration, one additional constraint (\ref{eq:master_2}) is added to the master problem. Hence, the optimum of the master problem is non-increasing with number of iterations. 

The primal problem gives a solution which is a lower bound to the optimum of original problem $\textbf{P}_1$ as it provides solution for fixed $\vec{I}_H$. In each iteration, the lower bound is set equal to the maximum of the lower bounds obtained in current and previous iteration.

In the $l$th iteration the primal problem is solved for the solution obtained by master problem in $(l-1)$th iteration. Then, for the obtained solution of the primal problem in iteration $l$, we solve the master problem. This process continues and due to non-increasing (non-decreasing) nature of the upper bound (lower bound), optimal solution can be obtained the GBD algorithm converges in finite number of iterations \cite{GBD}.

The primal problem is convex and can be solved in polynomial time. However, the master problem is NP-hard as it is an integer programming problem. However, GBD can be solved efficiently using any commercial optimization software such as MOSEK \cite{mosek}. The GBD algorithm is summarized in Algorithm \ref{algo:GBD}, where $\mathcal{S}$ is a set of constraint \eqref{eq:master_2} in which an additional
constraint is added in each iteration.
\begin{algorithm}
\caption{GBD algorithm}
\label{algo:GBD}
\begin{algorithmic}
\State \textbf{Initialization:} Initialize $\vec{I}_H^{(0)}$ randomly, convergence parameter $\epsilon$. Set $\mathcal{S}\leftarrow\emptyset$ and $j\leftarrow1$.
\State Set flag$\leftarrow1$
\While{flag$\neq0$}
\State Solve the primal problem (\ref{eq:primal_1}) and obtain

$\vec{p}_s^{*},\theta^*,\vec{\pmb{\lambda}}^*,\vec{\pmb{\gamma}}^*,\vec{\pmb{\delta}}^*,\mu^*$ and lower bound L$_{(j)}$
\State $\mathcal{S}\leftarrow \mathcal{S}\cup\{j\}$
\State solve master problem (\ref{eq:master_1}) and obtain $\vec{I}_H^{(j)*}$ and the upper bound U$_{(j)}$.
\If{$\vert$U$_{(j)}$-L$_{(j)}\vert\leq\epsilon$}
\State flag$\leftarrow0$
\EndIf
\State Set $j\leftarrow j+1$
\EndWhile \\
\Return $\vec{p}_s$ and $\vec{I}_H$
\end{algorithmic}
\end{algorithm}
\begin{figure}[!ht]
	\vspace{-3mm}
\centering
\includegraphics[width=0.83\linewidth]{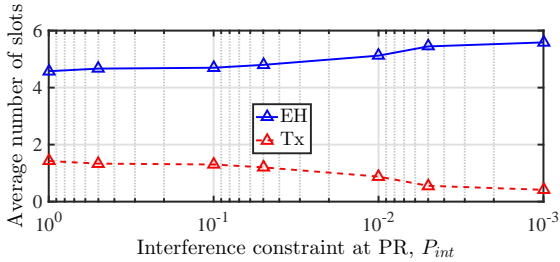}
  \caption{Average EH time and Tx time in versus interference threshold at PR ($M=6, N=10, E_0=2$, and $\alpha=0.9$).}
\label{fig:avg_EH_time}
\vspace{-2mm}
\end{figure}

\begin{figure}[!ht]
	\centering
	\includegraphics[width=0.83\linewidth]{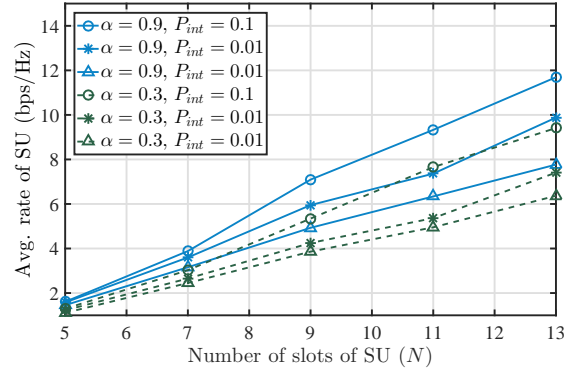} 
	\caption{Average achievable rate of ST versus the number of slots of SU ($E_0=2$ J).}
	\label{fig:B_s_vs_N_optimal}
\end{figure}
\begin{figure}[!ht]
	\centering
	\includegraphics[width=0.83\linewidth]{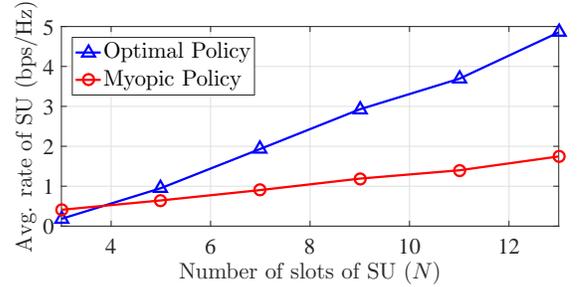}
	\caption{Average achievable rate of ST under the optimal and myopic policies ($E_0=0$, $P_{int}=0.1$ and $\alpha=0.3$).}
	\label{fig:comparison_myopic_optimal}
\end{figure}
\begin{figure}[!ht]
\centering
  \includegraphics[width=0.83\linewidth]{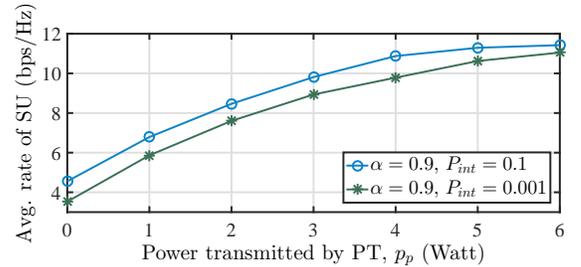}
  \caption{Average achievable rate of ST versus power transmitted by PT, $p_p$ ($E_0=2$J)}
\label{fig:throughput_vs_p_p}
\end{figure}
\section{Results}
\indent We study the performance of the optimal \emph{harvest-or-transmit} strategy in this section. We assume quasi static i.i.d. Rayleigh distributed channel links with variances $\sigma^2_{pp}=\sigma^2_{ps}=\sigma^2_{sp}=\sigma^2_{ss}=0.1$ and $\sigma^2=0.1$.
\subsection{Effect of $P_{int}$}
In Fig. \ref{fig:avg_EH_time}, the effect of $P_{int}$ on the average EH and average Tx time for the first $M$ slots is shown. After $M$ slots, the PU becomes silent and the SU can not harvest RF energy from it. We assume that the PT transmits with power $p_p=1$ W in all the $M$ slots and initial energy in the battery $E_0=2$ J. It is evident that as $P_{int}$ decreases, the average EH time increases and average Tx time decreases (the average Tx time is the duration as long as PT remains active). When $P_{int}$ approaches zero, average EH-time approaches $M$ and average Tx-time approaches 0, i.e., the ST harvests energy as long as PT is active.

Fig. \ref{fig:B_s_vs_N_optimal} shows the average achievable rate of ST under the optimal policy for different energy harvesting efficiency and different interference constraints at PR. The average is obtained over different channel realizations. For the simulation purpose, the number of primary slots, $M$ is assumed to be $N-2$. From the Fig. \ref{fig:B_s_vs_N_optimal}, it is evident that as the interference constraint at the primary receiver loosens, i.e., $P_{int}$ increases, and thus secondary transmitter is able to transmit with higher power, which results in higher achievable rate. When the interference constraint becomes too stringent, the secondary transmitter can not transmit as long as the PU is present. So, in this case, it harvests in first $M$ slots and transmits in remaining $N-M$ slots with total available energy of $E_0+M\alpha p_p$ and since ST does not get enough time to transmit, its achievable rate decreases as $P_{int}$ decreases.

Fig. \ref{fig:comparison_myopic_optimal} shows the comparison between the optimal and the myopic policy proposed in \cite{underlay_3}. Since each slot at the ST is dedicated for either harvesting energy from the PU or transmitting its data, for smaller number of slots, the ST may not use the available slots efficiently for its transmission and therefore, the average achievable rate in our policy is less than that of in \cite{underlay_3}. However, as the number of secondary slots $N$ increases, the proposed policy outperforms the myopic policy \cite{underlay_3} as shown in Fig. \ref{fig:comparison_myopic_optimal}. This is because in our policy, the ST takes the future channel gains into account and optimizes its transmit power over all the slots jointly.

Fig. \ref{fig:throughput_vs_p_p} shows the average achievable rate versus the power transmitted by PT for fixed $\alpha$ and different values of $P_{int}$ averaged over different channel realizations. For simulation purpose we assumed $M$ and $N$ to be 6 and 10 respectively and $\alpha=0.9$. From the figure, it is inferred that the rate increases with $p_p$ because with increasing $p_p$, ST harvests more energy in each harvesting slot and can transmit with higher power.

\vspace{0mm}
\begin{figure}[!t]
\centering
\includegraphics[width=0.83\linewidth]{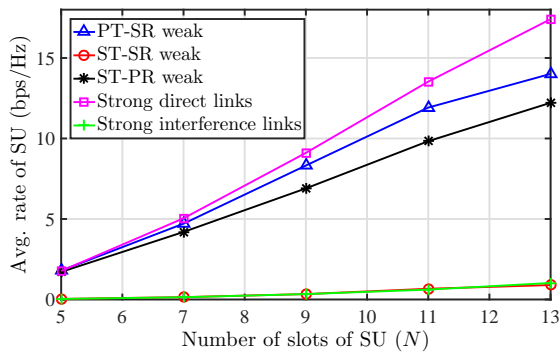}
\caption{Effects of different channel conditions on average achievable rate of ST ($E_0=2$ J and $P_{int}=0.1$). }
\label{fig:E_0_vs_time}
\end{figure}
\subsection{Effects of different channel conditions}
Fig. \ref{fig:E_0_vs_time} shows the effect of different channel conditions on achievable rate. For simulation purpose, we assume the variance of weak links to be 0.01 and variance of strong links to be 0.1, $E_0=2$ J and $\alpha=0.9$. From the figure, it can be observed that when direct links are strong, the achievable rate is maximum as due to weak interference links, ST causes less interference to PR and receives less interference from PT. This allows ST to increase its transmission power which results in higher rate. The weak ST-SR and strong interference link case performs worst in all the scenarios because in both of these cases, SR receives more interference from PT and ST causes more interference to PR due to which ST can not transmit with higher power. Also, when ST-PR and PT-SR links are weak, performance degrades due to similar reasons.
\section{Conclusions}
\indent We obtained the optimal \textit{harvest-or-transmit} policy of an underlay EH-CRN using GBD algorithm and studied the effects of different system parameters. We observed that the optimal EH (Tx) time increases (decreases) as $P_{int}$ decreases. Also, we analyzed the effects of $P_{int}$ on average achievable rate and observed that it reduces as $P_{int}$ decreases. The effect of various channel conditions on average achievable rate has also been studied. In addition, we showed that the proposed policy outperforms the myopic policy proposed in the literature.


\bibliographystyle{ieeetr} 
\bibliography{references}

\end{document}